\RequirePackage[2020-02-02]{latexrelease}
\documentclass[eqsecnum,superscriptaddress,noshowpacs,showkeys,nofootinbib,aps]{revtex4}
\usepackage{epsfig}
\usepackage{amsmath,latexsym}
\usepackage{multirow}

\usepackage{xcolor}
\usepackage{ulem}

\renewcommand{\theequation}{\arabic{equation}}
\def\be{\begin{equation}}
\def\ee{\end{equation}}
\def\bea{\begin{eqnarray}}
\def\eea{\end{eqnarray}}

\begin{document}

\title{GUP corrected entropy of the Schwarzschild black hole in holographic massive gravity}
\author{Soon-Tae Hong}
\email{galaxy.mass@gmail.com}
 \affiliation{Center for Quantum Spacetime and
 \\ Department of Physics, Sogang University, Seoul 04107, Korea}

\author{Yong-Wan Kim}
\email{ywkim65@gmail.com}
 \affiliation{Department of Physics and
 \\ Research Institute of Physics and Chemistry, Jeonbuk National University, Jeonju 54896, Korea}

\author{Young-Jai Park}
 \affiliation{Center for Quantum Spacetime and
 \\ Department of Physics, Sogang University, Seoul 04107, Korea}

\date{\today}

\begin{abstract}
We obtain the statistical entropy of a scalar field on the
Schwarzschild black hole in holographic massive gravity by
considering corrections on the density of quantum states to all
orders in the Planck length from a generalized uncertainty
principle (GUP). As a result, we find not only the generalized
Bekenstein-Hawking entropy depending on holographically massive
gravitons without any artificial cutoff, but also new additional
correction terms, which are proportional to surface gravity.
Moreover, we also observe that all order GUP corrected entropy is
improved to have smaller GUP parameter $\lambda$ than the previous
results.
\end{abstract}



\maketitle

\section{introduction}
\setcounter{equation}{0}
\renewcommand{\theequation}{\arabic{section}.\arabic{equation}}

Einstein's theory of general relativity (GR) is a theory of a
massless spin-2 graviton, which has been successfully tested to
date as the description of the force of gravity. However, quantum
gravity phenomenology \cite{AmelinoCamelia:2008qg} that focuses on
modifications of the existing theory at extreme limits has pushed
forward to search for alternatives to GR. One of them is to
introduce a massive graviton to GR. In 1930s, by extending GR with
a quadratic mass term, Fierz and Pauli developed a massive spin-2
theory \cite{Fierz:1939ix}. It was later known to suffer from the
Boulware-Deser ghost problem \cite{Boulware:1973my} and the van
Dam, Veltman and Zakharov (vDVZ) discontinuity
\cite{vanDam:1970vg,Zakharov:1970cc} in the massless graviton
limit. The vDVZ discontinuity was cured by the Vainshtein
mechanism \cite{Vainshtein:1972sx} due to certain low scale
strongly coupled interactions. Moreover, de Rham, Gabadadze and
Trolley (dRGT) \cite{deRham:2010ik,deRham:2010kj} successfully
obtained a ghost free massive gravity, which has nonlinearly
interacting mass terms constructed from the metric coupled with a
symmetric reference metric tensor. This was confirmed by a
Hamiltonian analysis of the untruncated theory
\cite{Hassan:2011hr,Hassan:2011tf} and other works
\cite{Kluson:2011qe,Kluson:2011rt,Kluson:2012gz,Comelli:2012vz,Golovnev:2011aa,Deffayet:2012nr}.
Furthermore, Vegh \cite{Vegh:2013sk} introduced a nonlinear
massive gravity with a special singular reference metric which
keeps the diffeomorphism symmetry for coordinates ($t,r$) intact
but breaks it in angular directions so that gravitons acquire the
mass because of a broken momentum conservation
\cite{Davison:2013jba,Blake:2013bqa,Blake:2013owa}. Since then,
this Vegh's type of massive gravity, called holographic massive
gravity, has been extensively exploited to investigate many black
hole models
\cite{Cai:2014znn,Adams:2014vza,Hendi:2015pda,Hu:2016hpm,Zou:2016sab,
Hendi:2017fxp,Tannukij:2017jtn,Hendi:2017bys,Hendi:2018xuy,Chabab:2019mlu,Hong:2018spz}.
Very recently, we have investigated the tidal effects in the
Schwarzschild black hole in holographic massive gravity (SBHHMG),
showing that massive gravitons effectively affect the angular
component of the tidal force, while the radial component remains
the same as in massless gravity \cite{Hong:2019zsi}.

On the other hand, quantum gravity phenomenology predicts the
possible existence of a minimal length on the smallest scale [33].
This implies the modification of the Heisenberg uncertainty
principle (HUP) in quantum mechanics to a generalized uncertainty
principle (GUP)
\cite{Kempf:1994su,Garay:1994en,Scardigli:1999jh,KalyanaRama:2001xd,
Chang:2001bm,Hossenfelder:2012jw}, which paves the way for deeper
understanding on black hole thermodynamics
\cite{Adler:2001vs,Scardigli:2003kr,Setare:2004sr,Medved:2004yu,Ling:2005bq,Myung:2006qr,Kim:2016qtp,Feng:2020lai}
including the final stage of the Hawking radiation. As is well
known, studies on black holes in terms of thermodynamics started
with the discovery of the Bekenstein-Hawking entropy
\cite{Bekenstein:1972tm,Bekenstein:1973ur,Bekenstein:1974ax,
Hawking:1974rv,Hawking:1974sw} proportional to the surface area of
a black hole at the event horizon. In order to provide a
microphysical explanation to the Bekenstein-Hawking entropy, 't
Hooft developed the statistical method of finding the black hole's
entropy by introducing a scalar field propagating just outside the
event horizon \cite{tHooft:1984kcu}. After the pioneering work of
't Hooft,  in the last several decades, a lot of authors have
studied statistical properties of various types of black holes
\cite{Mann:1990fk,Ghosh:1994wb,Demers:1995dq,Cai:1996js,
Kim:1996bp,Kim:1996eg,Ho:1997fg,Mukohyama:1998rf,Jing:1999bw,Li:2000rk,
Winstanley:2000in,Kim:2001kpa,Medved:2001zw,
Gao:2002ed,Sun:2004ct,Kenmoku:2005zh,Sarkar:2007uz,Singleton:2010gz,
Eune:2012mv,Lenz:2014aea,Eune:2014tea,Kamali:2016agu,Kamali:2018,Vagenas:2019wzd,Cuadros-Melgar:2020shz}.
One of main characteristics in his so-called brick wall method is
to introduce an cutoff which removes ultraviolet divergences due
to the infinite blue shift at the event horizon. Later, it was
shown that the cutoff can be effectively replaced by a minimal
length to the order of the Planck length which is derived from a
GUP \cite{Li:2002xb}. Making use of these ideas, the authors in
Refs. \cite{Zhao:2003eu,Liu:2004xh,Kim:2006rx,Yoon:2007aj} have
calculated the statistical entropy of black holes to leading order
in the Planck length. The ultraviolet divergences of the just
vicinity near the horizon in the usual brick wall method is
drastically solved by the newly modified equation of the density
states motivated by GUPs
\cite{Kempf:1994su,Garay:1994en,Scardigli:1999jh,KalyanaRama:2001xd,
Chang:2001bm,Hossenfelder:2012jw}. Since the GUP up to leading
order correction in the Planck length is not enough because the
wave vector $k$ does not satisfy the asymptotic property in the
modified dispersion relation
\cite{Hossenfelder:2006cw,Hossenfelder:2005ed}, Nouicer has
further developed the GUP effect to all orders in the Planck
length \cite{Nouicer:2007jg,Nouicer:2007cw} After his work,
according to this approach, we have obtained the desired
Bekenstein-Hawking entropy to all orders in the Planck length
units without any artificial cutoff and little mass approximation
for the case of the Schwarzschild black hole \cite{Kim:2007if}.
However, all of these works have been concentrated on black holes
in GR, and it has been rarely studied in alternatives to GR such
as massive gravity, as far as we know. Therefore, it would be
interesting to extend them to massive gravity.

In this paper, we calculate statistical entropy of a scalar field
on the Schwarzschild black hole in holographic massive gravity by
imposing GUP effect additionally to all orders in the Planck
length. In Sec. II, we briefly recapitulate the formalism for all
order corrections of GUP and its relation to existence of minimal
length. In Sec. III, we find the solution of the SBHHMG for
self-consistency and discuss Hawking temperatures according to
mass parameters. In Sec. IV, we consider a scalar field
propagating on the SBHHMG , and calculate the generalized entropy
to all orders in the Planck length by counting modified density
states in the presence of massive gravitons. Finally, summary and
discussion are drawn in Sec. V.

\section{The formalism of all order corrections of GUP and minimal length}

Quantum gravity phenomenology has been tackled with effective
models which incorporate a minimal length as a natural ultraviolet
cutoff \cite{Hossenfelder:2006cw,Hossenfelder:2005ed}. Such a
minimal length leads to deformed Heisenberg algebras
\cite{Kempf:1994su,Garay:1994en,Scardigli:1999jh,KalyanaRama:2001xd,Adler:2001vs}
which show a GUP. For a particle with the momentum $p$ and the
wave vector $k$ having a nonlinear relation $p=f(k)$, the
commutator between two operators $\hat{x}$ and $\hat{p}$ can be
generalized to
\begin{equation}\label{xp}
 [\hat{x}, \hat{p}] = i \frac{\partial p}{\partial k}
 ~\Leftrightarrow
 \Delta x \Delta p \geq \frac{1}{2} \left|~\left< \frac{\partial p}{\partial k} \right> ~\right|
\end{equation}
at the quantum mechanical level
\cite{Hossenfelder:2006cw,Hossenfelder:2005ed}. Without loss of
generality, in the following, let us restrict ourselves to the
isotropic case in one space-like dimension.

As a nonlinear dispersion relation, Kempf et al.
\cite{Kempf:1994su,Garay:1994en,Scardigli:1999jh,KalyanaRama:2001xd,Adler:2001vs}
have considered the following relation
\begin{equation}\label{KMM}
\frac{\partial p}{\partial k}=1+\lambda p^2,
\end{equation}
which came in the context of perturbative string theory. The GUP
parameter $\lambda$ is of order of the Planck length $l^{-2}_p$.
By solving the GUP in Eq. (\ref{xp}), one can find that it
exhibits the features of UV/IR correspondence such that as $\Delta
p$ is large, $\Delta x$ is proportional to $\Delta p$. More
importantly, it provides us the existence of the minimal length as
$(\Delta x)^{\rm Kempf}_{\rm min}=2\sqrt{\lambda}$ below which
spacetime distances cannot be resolved.

On the other hand, the GUP of Kempf et al. was extended to all
orders in the Planck length
\cite{Smailagic:2003yb,Smailagic:2003rp,Nouicer:2007hz} as
\begin{equation}\label{Nouicer}
\frac{\partial p}{\partial k}=~e^{\lambda p^{2}}.
\end{equation}
For this case, the GUP in Eq. (\ref{xp}) for mirror symmetric
states as $\langle\hat{p}\rangle=0$ can be solved by the
multi-valued Lambert function \cite{Corless:1996zz}. In order to
have a real solution for $\Delta p$, it is required to satisfy the
following position uncertainty
\cite{Kim:2007if,Smailagic:2003yb,Smailagic:2003rp,Nouicer:2007hz}
\begin{equation}
\label{minlength}
 \Delta x\ge\sqrt{\frac{e\lambda}{2}}.
\end{equation}
Therefore, one can readily rewrite the minimal length $(\Delta
x)_{\rm min}=\sqrt{e/8}(\Delta x)^{\rm Kempf}_{\rm min}$ for the
GUP to the all orders in the Planck length.

Note that this includes the leading order correction in the
relation (\ref{KMM}). However, since this correction only of the
GUP does not satisfy the property that the wave vector $k$
asymptotically reaches the cutoff in large energy region as in
Ref. \cite{Hossenfelder:2006cw,Hossenfelder:2005ed}, we will
consider the all order corrections in the Planck length in the
followings.

\section{Hawking temperature of the SBHHMG}

The (3+1)-dimensional SBHHMG
\cite{Vegh:2013sk,Davison:2013jba,Blake:2013bqa,Blake:2013owa,
Cai:2014znn,Adams:2014vza,Hendi:2015pda,Hu:2016hpm,Zou:2016sab,
Hendi:2017fxp,Tannukij:2017jtn,Hendi:2017bys,Hendi:2018xuy,
Chabab:2019mlu,Hong:2018spz,Hong:2019zsi} is described by the
action
 \be\label{mSch}
 S=\frac{1}{16\pi G}\int d^4x\sqrt{-g}\left[{\cal R}
   +{\tilde m}^2\sum_{a=1}^{4}c_a{\cal U}_{a}(g_{\mu\nu},f_{\mu\nu})\right],
 \ee
where ${\cal R}$ is the scalar curvature of the metric
$g_{\mu\nu}$, $\tilde{m}$ is a graviton mass\footnote{In this
paper, we shall call it massless when $\tilde{m}$ is zero.}, $c_a$
are constants, and ${\cal U}_a$ are symmetric polynomial
potentials of the eigenvalue of the matrix ${\cal
K}^\mu_\nu\equiv\sqrt{g^{\mu\alpha}f_{\alpha\nu}}$ as
 \bea
 {\cal U}_1 &=& [{\cal K}],\nonumber\\
 {\cal U}_2 &=& [{\cal K}]^2-[{\cal K}^2],\nonumber\\
 {\cal U}_3 &=& [{\cal K}]^3-3[{\cal K}][{\cal K}^2]+2[{\cal K}^3],\nonumber\\
 {\cal U}_4 &=& [{\cal K}]^4-6[{\cal K}^2][{\cal K}]^2+8[{\cal K}^3][{\cal K}]+3[{\cal K}^2]^2-6[{\cal K}^4].
 \eea
Here, the square root in ${\cal K}$ means
$(\sqrt{A})^\mu_\alpha(\sqrt{A})^\alpha_\nu=A^\mu_\nu$ and square
brackets denote the trace $[{\cal K}]={\cal K}^\mu_\mu$. Indices
are raised and lowered with the dynamical metric $g_{\mu\nu}$,
while the reference metric $f_{\mu\nu}$ is a non-dynamical, fixed
symmetric tensor which is introduced to construct nontrivial
interaction terms in holographic massive gravity.

Variation of the action (\ref{mSch}) with respect to the metric
$g_{\mu\nu}$ leads to the equations of motion given by
 \begin{eqnarray}
 \label{eom}
 {\cal R}_{\mu\nu}
 &-& \frac{1}{2}g_{\mu\nu}\left({\cal R}+\tilde{m}^2\sum^{4}_{a=1}c_a{\cal U}_a\right)\nonumber\\
 &+&\frac{1}{2}\tilde{m}^2\sum^{4}_{a=1}\left[ac_a{\cal U}_{a-1}{\cal K}_{\mu\nu}
 -a(a-1)c_a{\cal U}_{a-2}{\cal K}^2_{\mu\nu}
 +6(3a-8)c_a{\cal U}_{a-3}{\cal K}^3_{\mu\nu}
 -12(a-2)c_a{\cal U}_{a-4}{\cal K}^4_{\mu\nu}\right]=0.
 \end{eqnarray}
with ${\cal U}_{-a}=0$ and ${\cal U}_0=1$.

When one considers the spherically symmetric black hole solution
ansatz as
 \be\label{metric-mSch}
 ds^2=-f(r)dt^2+f^{-1}(r)dr^2+r^2(d\theta^2+\sin^2\theta d\phi^2)
 \ee
with the following degenerate reference metric
\cite{Vegh:2013sk,Davison:2013jba,Blake:2013bqa,Blake:2013owa,Cai:2014znn,Adams:2014vza,Hendi:2015pda,Hu:2016hpm,Zou:2016sab,
Hendi:2017fxp,Tannukij:2017jtn,Hendi:2017bys,Hendi:2018xuy,Chabab:2019mlu,Hong:2018spz}
 \be\label{fidmetric}
 f_{\mu\nu}={\rm diag}(0,0,c^2_0,c^2_0\sin^2\theta),
 \ee
one can find
 \be
 {\cal K}^\theta_\theta={\cal K}^\phi_\phi=\frac{c_0}{r}.
 \ee
Note that the choice of the reference metric in Eq.
(\ref{fidmetric}) preserves general covariance in ($t,r$) but not
in the angular directions. This gives the symmetric potentials as
 \be
 {\cal U}_1=\frac{2c_0}{r},~~~{\cal U}_2=\frac{2c^2_0}{r^2},~~~{\cal U}_3={\cal
 U}_4=0.
 \ee
It should be pointed out that there are no contributions from
$c_3$ and $c_4$ terms which appear in (4+1) and (5+1)-dimensional
spacetimes, respectively
\cite{Cai:2014znn,Hendi:2015pda,Hu:2016hpm,Zou:2016sab,Hendi:2017fxp,Hendi:2018xuy}.
Then, one can have the solution
 \be\label{lapsewol}
 f(r)=1-\frac{2m}{r}+2Rr+{\cal C}
 \ee
with $R=c_0c_1\tilde{m}^2/4$ and ${\cal C}=c^2_0c_2\tilde{m}^2$
\cite{Hong:2019zsi}, where $m$ is an integration constant related
to the mass of the black hole and  $c_0$ is a positive constant
\cite{Vegh:2013sk,Davison:2013jba,Blake:2013bqa,Blake:2013owa,Cai:2014znn,Adams:2014vza,Hendi:2015pda,Hu:2016hpm,Zou:2016sab,
Hendi:2017fxp,Tannukij:2017jtn,Hendi:2017bys,Hendi:2018xuy,Chabab:2019mlu,Hong:2018spz}.

Now, by solving $f(r)=0$, one can find the event horizons of the
SBHHMG as
\begin{equation}
 r_\pm=\frac{-(1+{\cal C})\pm\sqrt{16mR+(1+{\cal C})^2}}{4R}.
\end{equation}
The allowed event horizon can be classified according to the
relative signs of $R$ and ${\cal C}$ as shown in Table \ref{evth}.
We note that the event horizon $r_+$ of the SBHHMG is reduced to
$2m$ of the Schwarzschild black hole  in massless gravity as $R$
and ${\cal C}\rightarrow 0$. It is also appropriate to comment
that in Table \ref{evth}, $r_-$ is discarded since it is either
negative or imaginary in each ranges. Only in the special case of
${\cal C}>-1$ and $R=-\frac{(1+{\cal C})^2}{16m}$, the physical
solution of $r_-$ exists and coincides exactly with $r_+$. In
Table \ref{evth}, the abbreviation NA denotes that there is no
available event horizon in the specified range.
\begingroup
\setlength{\tabcolsep}{6pt} 
\renewcommand{\arraystretch}{1.8}
\begin{table}[ht!]
  \begin{center}
    \caption{Event horizons of the SBHHMG}
    \label{evth}
    \begin{tabular}{|c|cc|c|c|}
    \hline
            &  & ${\cal C}>-1$ &  ${\cal C}=-1$  & ${\cal C}<-1$ \\
    \hline
      $R>0$ & (I) & $r_+=\frac{1+{\cal C}}{4R}\left(-1+\sqrt{1+\frac{16mR}{(1+{\cal C})^2}}~\right)$     &  (II) $r_+=\sqrt{\frac{m}{R}}$~   &  (III) $r_+=-\frac{1+{\cal C}}{4R}\left(1+\sqrt{1+\frac{16mR}{(1+{\cal C})^2}}~\right)$  \\
      $R=0$ & (IV) & $r_+=\frac{2m}{1+{\cal C}}$      &    NA   & NA \\
      ~$-\frac{(1+{\cal C})^2}{16m}<R<0$~ & (V) & $r_+=\frac{1+{\cal C}}{4R}\left(-1+\sqrt{1+\frac{16mR}{(1+{\cal C})^2}}~\right)$        &    NA   &   NA \\
      $R=-\frac{(1+{\cal C})^2}{16m}$ & (VI) & $r_\pm=-\frac{1+{\cal C}}{4R}$        &    NA      &  NA \\
      $R<-\frac{(1+{\cal C})^2}{16m}$ & &  NA      &      NA   &     NA \\
    \hline
    \end{tabular}
  \end{center}
\end{table}
\endgroup

Moreover, from the solution (\ref{lapsewol}), one can find the
surface gravity~\cite{wald}
 \be\label{kh-msch}
 \kappa_+ =\frac{1+{\cal C}}{2r_+}+2R,
 \ee
and the Hawking temperature $T_H$ for the SBHHMG
\cite{Hong:2019zsi} as
 \be\label{HTmSch}
 T_H=\frac{1+{\cal C}}{4\pi r_+}+\frac{R}{\pi}.
 \ee

In Table \ref{THs}, we have summarized the Hawking temperatures in
the holographic massive gravity which behave differently according
to the relative signs of $R$ and ${\cal C}$. First of all, in the
case of (I) and (IV), the Hawking temperatures are depicted as the
one in massless gravity, proportional to $\frac{1}{2m}$ but
approach to a constant of $\frac{R}{\pi}$ as $m\rightarrow 0$. In
the case of (II) with ${\cal C}=-1$, the Hawking temperature is
given by a constant, and in the case of (III) with ${\cal C}<-1$,
the curve is flipped due to the negative sign of ${\cal C}<-1$ in
front of $\frac{1}{4\pi r_+}$ in Eq. (\ref{HTmSch}) and it becomes
$\frac{R}{2\pi}$ as $m\rightarrow 0$. In the case of (V), the
Hawking temperature is the same with (I), however, the range of
$m$ is limited to $m<-\frac{(1+{\cal C})^2}{16R}$ (note that
$R<0$). Finally, in the case of (VI), the Hawking temperature
vanishes.
\begingroup
\setlength{\tabcolsep}{1pt} 
\renewcommand{\arraystretch}{1.8}
\begin{table}[ht!]
  \begin{center}
    \caption{Hawking temperatures of the SBHHMG}
    \label{THs}
    \begin{tabular}{|c|cc|c|c|}
    \hline
            &  & ${\cal C}>-1$ &  ${\cal C}=-1$  & ${\cal C}<-1$ \\
    \hline
      $R>0$ & (I) & $T_H=\frac{R}{\pi}\left[1+\frac{(1+{\cal C})^2}{16mR}\left(1+\sqrt{1+\frac{16mR}{(1+{\cal C})^2}}~\right)\right]$     &  (II) $T_H=\frac{R}{\pi}$~   &  (III) $T_H=\frac{R}{\pi}\left[1+\frac{(1+{\cal C})^2}{16mR}\left(1-\sqrt{1+\frac{16mR}{(1+{\cal C})^2}}~\right)\right]$   \\
      $R=0$ & (IV) & $T_H=\frac{(1+{\cal C})^2}{8\pi m}$      &    NA   & NA \\
      ~$-\frac{(1+{\cal C})^2}{16m}<R<0$~ & (V) &  $T_H=\frac{R}{\pi}\left[1+\frac{(1+{\cal C})^2}{16mR}\left(1+\sqrt{1+\frac{16mR}{(1+{\cal C})^2}}~\right)\right]$        &    NA   &   NA \\
      $R=-\frac{(1+{\cal C})^2}{16m}$ & (VI) & $T_H=0$      &    NA      &  NA \\
      $R<-\frac{(1+{\cal C})^2}{16m}$ & &  NA      &      NA   &     NA \\
    \hline
    \end{tabular}
  \end{center}
\end{table}
\endgroup

\begin{figure}[t!]
   \centering
   \includegraphics{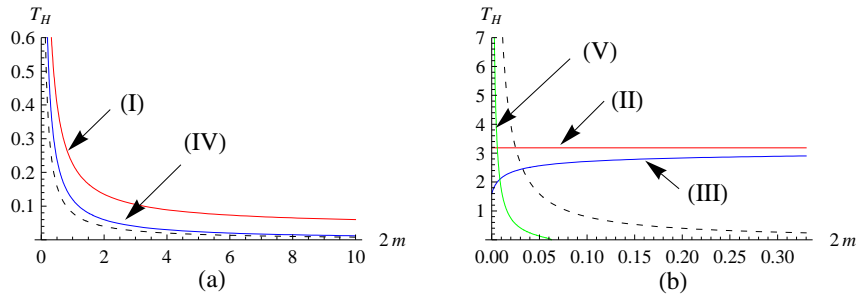}
\caption{Hawking temperatures of the SBHHMG, (a) for the case of
(I) with ${\cal C}=0.5$, $R=0.1~({\rm red})$ and (IV) with ${\cal
C}=0.5$, $R=0~({\rm blue})$, (b) for the case of (II) with ${\cal
C}=-1$, $R=10~({\rm red})$, (III) with ${\cal C}=-1.5$,
$R=10~({\rm blue})$, and (V) ${\cal C}=-0.5$, $R=-0.5~({\rm
green})$. The dashed black line is for the Hawking temperature in
massless gravity. Note that the horizontal axis is drawn in the
unit of $2m$ in order that the graphs are effectively compared
with the one of the Schwarzschild black hole in massless gravity.}
\label{fig1}
\end{figure}

These Hawking temperatures are drawn in Fig. \ref{fig1}. As shown
in Fig. \ref{fig1}(a), the Hawking temperatures in the holographic
massive gravity with ${\cal C}>-1$ and $R\ge 0$ are similar to the
one in massless gravity except approaching $\frac{R}{\pi}$ as $m
\rightarrow\infty$ for the case of (I). On the other hand, Fig.
\ref{fig1}(b) shows rather unexpected aspects of massive gravitons
in the Hawking temperatures. The case (II) gives us a constant
Hawking temperature due to the absence of the first term in Eq.
(\ref{HTmSch}), and the case (III) a reversed Hawking temperature
near $m=0$ due to the flip of sign in front of $\frac{1}{4\pi
r_+}$ while approaching $\frac{R}{\pi}$ as $m\rightarrow\infty$.
Finally, in the case of (V), the Hawking temperature decreases and
eventually vanishes at $m=-\frac{(1+{\cal C})^2}{16R}$. In the
followings, without any loss of generality, we will concentrate on
the cases of (I) and (IV) unless otherwise mentioned.

\section{Entropy of the SBHHMG to all orders in the Planck Length}

We begin by considering the SBHHMG found in the previous section
as
\begin{equation} \label{d4ss}
  ds^2 = - f(r) dt^2 + f^{-1}(r) dr^2
        + r^2 (d\theta^2+\sin^2\theta d\phi^2),
\end{equation}
where $f(r)=1-\frac{2m}{r}+2Rr+{\cal C}$. Then, let us consider a
free scalar field with a mass $\mu$ in the background described by
the solution (\ref{d4ss}), which satisfies the Klein-Gordon
equation given by
\begin{equation}
  \label{wveqn}
  (\Box - \mu^2)\Phi = \frac{1}{\sqrt{-g}}\partial_\mu (\sqrt{-g}g^{\mu\nu}\partial_\nu\Phi) - \mu^2\Phi =0.
\end{equation}
Substituting the ansatz of wave function $\Phi(t,r,\theta, \phi) =
e^{-i\omega t}\psi(r, {\theta}, \phi)$ into Eq. (\ref{wveqn}), we
find that the Klein-Gordon equation in the spherical coordinates
becomes
\begin{equation}
\label{rtheta0}
 \partial_{r}^2 \psi + \left( \frac{f'}{f} + \frac{2}{r}\right)\partial_{r} \psi
 + \frac{1}{f}\left({\frac{1}{r^2}}\left[\partial^2_\theta + {\rm cot}\theta
\partial_\theta + {\frac{1}{{\rm sin}^{2}\theta}}\partial^2_\phi \right] +
\frac{\omega^{2}}{f} - \mu^{2} \right)\psi = 0.
\end{equation}
Here, the prime denotes the derivative with respect to $r$. By
using the Wenzel-Kramers-Brillouin approximation
\cite{tHooft:1984kcu} with $\psi \sim e^{iS(r,\theta,\phi)}$ and
keeping the real parts, we have the following modified dispersion
relation
\begin{equation}
\label{wkb}
 p_\mu p^\mu = - \frac{\omega^2}{f} + f{p_{r}}^{2} + \frac{p^2_\theta}{r^2}
   + \frac{p_{\phi}^2}{r^2 {\rm sin}^2 \theta}=-\mu^2
\end{equation}
where $p_{r} = \frac{\partial S}{\partial r}$, $p_{\theta} =
\frac{\partial S}{\partial \theta}$ and  $p_{\phi} =
\frac{\partial S}{\partial \phi}$. Furthermore, we also obtain the
square module of momentum as follows
\begin{equation}
\label{smom}  p^{2}\equiv g^{rr}{p_{r}}^{2} + g^{\theta
\theta}{p_{\theta}}^{2}+ g^{\phi\phi}{p_{\phi}}^{2} =
\frac{\omega^{2}}{f} - \mu^{2}.
\end{equation}
Then, the volume in the momentum phase space is given by
\begin{eqnarray}
V_{p}(r,\theta)= \int dp_{r}dp_{\theta}dp_{\phi} =\frac{4\pi}{3}
\frac{r^2 {\rm sin}\theta}{\sqrt{f}}
\left(\frac{\omega^2}{f}-\mu^2 \right)^{\frac{3}{2}}
\end{eqnarray}
with the condition $\omega\geq\mu\sqrt{f}$.

Now, let us calculate the statistical entropy of the scalar field
on the SBHHMG by imposing the additional GUP effect to all orders
in the Planck length units. When the gravity is turned on, the
number of quantum states in a volume element in phase cell space
based on the GUP in (3+1)-dimensions is given by
\begin{equation}
\label{dn} dn_A = \frac{d^3 x d^3 p}{(2\pi)^3}e^{-\lambda p^2} ,
\end{equation}
where $p^2$ is given in Eq. (\ref{smom}) and one quantum state
corresponding to a cell of volume is changed from $(2\pi)^3 $ into
$(2\pi)^3 e^{\lambda p^2}$ in the phase space
\cite{Li:2002xb,Zhao:2003eu,Liu:2004xh,Kim:2006rx}. Here, the
subscript $A$ denotes the quantity for all orders in the Planck
length. Note that in the limit of $\lambda\rightarrow 0$, we have
the number of quantum states with HUP \cite{tHooft:1984kcu}. From
Eqs. (\ref{smom}) and (\ref{dn}), the number of quantum states
related to the radial mode with energy less than $\omega$ is given
by
\begin{eqnarray}
\label{Tnqs} n_A(\omega) = \frac{2}{3\pi}\int_{r_+} dr
\frac{r^2}{\sqrt{f}} \left(\frac{{\omega}^2}{f}-
\mu^{2}\right)^{\frac{3}{2}} e^{-\lambda (\frac{{\omega}^2}{f}-
\mu^{2})}.
\end{eqnarray}
It is interesting to note that $n_A(\omega)$ is convergent at the
horizon without any artificial cutoff because of the existence of
the suppressing exponential $\lambda$-term induced from the GUP.

For the bosonic case, the free energy of a thermal ensemble of
scalar fields at inverse temperature $\beta$ is given by
\begin{eqnarray}
\label{TfreeE}
 F_{A}&=& \frac{1}{\beta}\sum_K \ln \left( 1 - e^{-\beta \omega_K}
 \right)\nonumber\\
     &=& - \frac{2}{3\pi} \int_{r_+} dr \frac{r^2}{\sqrt{f}}
   \int^{\infty}_{\mu\sqrt{f}} d\omega~
    \frac{\left(\frac{{\omega}^2}{f}- \mu^{2}\right)^{\frac{3}{2}}}{e^{\beta \omega} -1}e^{- \lambda (\frac{{\omega}^2}{f}- \mu^{2})}.
\end{eqnarray}
Here, we have considered the continuum limit, integrated it by
parts and used the number of quantum states (\ref{Tnqs}).

Now, we are only interested in the contribution from just the
vicinity near the event horizon in the range of $(r_+, r_+ +
\epsilon)$ where $\epsilon$ is the brick wall cutoff used to
remove ultraviolet divergences. Since $f \rightarrow 0$ near the
event horizon, $\frac{{\omega}^2}{f}-\mu^{2}$ becomes
$\frac{{\omega}^2}{f}$ so that we do not need to require the
little mass approximation. Then, the free energy can be rewritten
as
 \begin{equation}
 \label{TfreeEf0}
 F_{A} = - \frac{2}{3\pi} \int^{r_++\epsilon}_{r_+} dr
 \frac{r^2}{f^2} \int^{\infty}_{0} d\omega
 \frac{{\omega}^3}{e^{\beta \omega} -1} e^{- \lambda {\frac{{\omega}^2}{f}}}.
 \end{equation}

On the other hand, the minimal length (\ref{minlength}) is related
to a proper distance of $\epsilon$ order as
\begin{eqnarray}
\label{invariant} (\Delta x)_{\rm min}=
                 \sqrt{\frac{e\lambda}{2}}
                 \equiv
             \int_{r_+}^{r_++\epsilon} \frac{dr}{\sqrt{f(r)}}
                \approx \int_{r_+}^{r_+ +\epsilon}
                          \frac{dr}{\sqrt{2\kappa_+(r-r_+)}}
                 = \sqrt{\frac{2\epsilon}{\kappa_+}},
\end{eqnarray}
where the metric function $f(r)$ is expanded near the event
horizon as
\begin{equation}
\label{Tay}
 f(r) \approx f(r_+) +2\kappa_+
 (r-r_+) + {\cal O}\left( (r-r_+)^2 \right),
\end{equation}
and in the second term $\frac{1}{2}\frac{df}{dr}|_{r=r_+}$  is
replaced by the surface gravity $\kappa_+$ (\ref{kh-msch}). Thus,
one can use the minimal length as a natural ultraviolet cutoff,
which shall provide a convergent entropy integral.

Then, from $F_A$ in Eq. (\ref{TfreeEf0}), one can find the entropy
as
\begin{eqnarray}
\label{Aentropy0}
S_{A} &=& \beta^2 \frac{\partial F_A}{\partial \beta}\mid_{\beta=\beta_+} \nonumber \\
      &=& \frac{\beta^2_+}{6\pi}
          \int^{r_+ +\epsilon}_{r_+} dr \frac{r^2}{f^2}
          \int^{\infty}_{0} d\omega\frac{{\omega}^4 }{\sinh^2(\frac{\beta_+}{2}\omega)}
           e^{- \lambda {\frac{{\omega}^2}{f}}}.
\end{eqnarray}
By defining $x\equiv \sqrt{\lambda}\omega$, this can be recast by
\begin{eqnarray}
\label{Aentropyx}
  S_{A} = \frac{\beta^2_+}{6\pi\lambda^2\sqrt{\lambda}} \int^{\infty}_{0}
          dx \frac{{x}^4}{\sinh^2(\frac{\beta_+}{2\sqrt{\lambda}}x)}\Lambda(x,\epsilon),
\end{eqnarray}
where
\begin{eqnarray}
\label{RfreeEf}
 \Lambda(x,\epsilon)  \equiv  \int^{r_+ + \epsilon}_{r_+} dr~\frac{r^2}{f^2}~e^{-\frac{x^2}{f}}.
\end{eqnarray}
Now, in order to find a final expression of entropy, one needs to
perform proper integrations of Eqs. (\ref{Aentropyx}) and
(\ref{RfreeEf}), which are depending on both metric functions and
types of the GUP. In \cite{Kim:2007if}, we had calculated this by
expanding all the functions only to the leading order near the
horizon. However, in this work, we would integrate Eqs.
(\ref{Aentropyx}) and (\ref{RfreeEf}) by the substitution method,
which will give us next order correction terms to the entropy.
First of all, near the horizon, we expand Eq. (\ref{RfreeEf}) as
\begin{equation}
\label{RfreeEf1}
 \Lambda(x,\epsilon) \approx  \int^{r_+ + \epsilon}_{r_+} dr~
 \frac{r^2}{[2\kappa_+(r-r_+)]^2}~e^{-\frac{x^2}{2\kappa_+(r-r_+)}}.
\end{equation}
This can be integrated by the substitution of
$t=x^2/2\kappa_+(r-r_+)$ in an exact and closed form as
\begin{eqnarray}
 \Lambda(x,\epsilon)&=&\frac{1}{2\kappa_+ x^2}\int^{\infty}_{\frac{x^2}{2\kappa_+\epsilon}} dt
          \left(r^2_+ +\frac{r_+x^2}{\kappa_+ t}+\frac{x^4}{4\kappa^2_+ t^2}\right)e^{-t}\nonumber\\
                    &=& \frac{r^2_+}{2\kappa_+ x^2}\Gamma(1,\frac{x^2}{2\kappa_+\epsilon})
                        +\frac{r_+}{2\kappa^2_+}\Gamma(0,\frac{x^2}{2\kappa_+\epsilon})
                        +\frac{x^2}{8\kappa^3_+}\Gamma(-1,\frac{x^2}{2\kappa_+\epsilon}),
\end{eqnarray}
where we have used the incomplete Gamma function given by
\begin{equation}
\Gamma(a,z)=\int^\infty_z dt~ t^{a-1}e^{-t}.
\end{equation}
The incomplete Gamma function was also used in efficiently finding
the entropy for the noncommutative acoustic black hole
\cite{Anacleto:2014apa}.

Then, all order GUP corrected entropy can be written as
\begin{eqnarray}
\label{Aentropy1}
 S_A = \frac{\beta^2_+ r^2_+}{12\pi \lambda^2\sqrt{\lambda}\kappa_+}
      \int^{\infty}_{0} dx \frac{{x}^2\Gamma(1,\frac{x^2}{2\kappa_+\epsilon})}{\sinh^2(\frac{\beta_+ x}{2\sqrt{\lambda}})}
      +\frac{\beta^2_+ r_+}{12\pi \lambda^2\sqrt{\lambda}\kappa^2_+}
      \int^{\infty}_{0} dx \frac{{x}^4\Gamma(0,\frac{x^2}{2\kappa_+\epsilon})}{\sinh^2(\frac{\beta_+ x}{2\sqrt{\lambda}})}
      +\frac{\beta^2_+}{48\pi \lambda^2\sqrt{\lambda}\kappa^3_+}
      \int^{\infty}_{0} dx
      \frac{{x}^6\Gamma(-1,\frac{x^2}{2\kappa_+\epsilon})}{\sinh^2(\frac{\beta_+
      x}{2\sqrt{\lambda}})}.\nonumber\\
\end{eqnarray}
Now, by redefining $y\equiv \frac{\beta_+ x}{2\sqrt{\lambda}}$ and
making use of the minimum length (\ref{invariant}) with $\beta_+
\kappa_+=2\pi$, we have
\begin{eqnarray}
\label{Fentropy1}
 S_A = \frac{r^2_+}{3\pi^2 \lambda}
      \int^{\infty}_{0} dy \frac{{y}^2\Gamma(1,\frac{2y^2}{\pi^2 e})}{\sinh^2y}
      +\frac{r_+\kappa_+}{3\pi^4}
      \int^{\infty}_{0} dy \frac{{y}^4\Gamma(0,\frac{2y^2}{\pi^2 e})}{\sinh^2y}
      +\frac{\lambda\kappa^2_+}{12\pi^6}
      \int^{\infty}_{0} dy \frac{{y}^6\Gamma(-1,\frac{2y^2}{\pi^2 e})}{\sinh^2y}.
\end{eqnarray}
The integrals can be numerically integrated as
\begin{eqnarray}
\label{delta1}
 \delta_1 &\equiv& \int^{\infty}_{0} dy \frac{{y}^2\Gamma(1,\frac{2y^2}{\pi^2 e})}{\sinh^2y}
                   \approx 1.4509,\\
 \delta_2 &\equiv& \int^{\infty}_{0} dy \frac{{y}^4\Gamma(0,\frac{2y^2}{\pi^2 e})}{\sinh^2y}
                   \approx 3.0709,\\
 \delta_3 &\equiv& \int^{\infty}_{0} dy \frac{{y}^6\Gamma(-1,\frac{2y^2}{\pi^2 e})}{\sinh^2y}
                   \approx 18.4609
\end{eqnarray}
so that the entropy can be obtained as
\begin{eqnarray}
\label{F1entropy1}
 S_A = \frac{\delta_1}{3\pi^3}\lambda^{-1}\left(\frac{A}{4}\right)
      +\frac{\delta_2}{3\pi^4}r_+\kappa_+
      +\frac{\delta_3}{12\pi^6}\lambda\kappa^2_+.
\end{eqnarray}
Here, $A=4\pi r^2_+$ is the surface area at the event horizon of
the SBHHMG. When we choose the GUP parameter $\lambda$ as
\begin{equation}
 \lambda=\frac{\delta_1}{3\pi^3}\approx 0.0156,
\end{equation}
we can finally obtain the entropy as
\begin{eqnarray}
\label{F2entropy1}
 S_A =\frac{A}{4}+\alpha r_+\kappa_++\beta\lambda\kappa^2_+,
\end{eqnarray}
where
\begin{equation}\label{values}
 \alpha=\frac{\delta_2}{3\pi^4}\approx 0.0105 ,
 ~~~\beta\lambda=\frac{\delta_1\delta_3}{36\pi^9}\approx 2.4959\times 10^{-5}.
\end{equation}
As a result, we have finally obtained the entropy satisfying the
area law with correction terms, which are proportional to the
surface gravity. It is important to comment that massive graviton
effect is included in the event horizon $r_+$ through $R$ and
$\cal C$.

It seems appropriate to comment on the GUP parameter $\lambda$ and
minimum length at this stage. As seen in Table
\ref{gup-parameters}, the GUP parameter $\lambda$ for the all
order GUP corrected $S_A$ is the smallest among other parameters
which are for the leading order GUP corrected $S_L$ based on
(\ref{KMM}) and the all order GUP corrected $S^{\rm old}_A$ with
the leading order approximation \cite{Kim:2007if}. On the other
hand, the all order GUP corrected $S_A$ has the smallest minimal
length and thinnest cutoff $\epsilon$. Note that in the all order
GUP corrected $S_A$ the first term is the same with the leading
order approximation in \cite{Kim:2007if}. However, it also
includes next higher order contributions expressed in Eq.
(\ref{delta1}). Moreover, by considering the integration using the
substitution method, we have obtained the all order GUP corrected
entropy with next orders of correction.
\begingroup
\setlength{\tabcolsep}{1pt} 
\renewcommand{\arraystretch}{1.8}
\begin{table}[ht!]
  \begin{center}
    \caption{Entropy, GUP parameter and minimal length. Note that the ellipses in the entropy are for next order terms and
    $\kappa_H$  in the last row is for the surface gravity in massless gravity.}
    \label{gup-parameters}
    \begin{tabular}{|c|c|c|c|c|}
    \hline
            &  entropy & ~ GUP parameter~ & ~ minimal length~& ~$\epsilon$ ~ \\
    \hline
      all order GUP correction in SBHHMG  & \, $S_A = \frac{\delta_1}{3\pi^3\lambda}\left(\frac{A}{4}\right)+\cdot\cdot\cdot$     & \, $\lambda=\frac{\delta_1}{3\pi^3}=0.0156$ &  $(\Delta x)_{\rm min}=\sqrt{e\lambda/2}=0.1456$ ~& $\frac{e}{4}\lambda\kappa_+$\\
      ~leading order GUP correction in SBHHMG & \,  $S_L = \frac{\zeta(3)}{3\pi\lambda}\left(\frac{A}{4}\right)$         & \, $\lambda=\frac{\zeta(3)}{3\pi}=0.1275$ & $(\Delta x)^{\rm Kemph}_{\rm min}=2\sqrt{\lambda}=0.4163$~& $2\lambda\kappa_+$\\
      ~all order GUP correction in massless gravity  & \, $S^{\rm old}_A = \frac{e^2\zeta(3)}{2\pi\lambda}\left(\frac{A}{4}\right)+\cdot\cdot\cdot$         & \,$\lambda=\frac{e^2\zeta(3)}{2\pi}=1.4136$~ & $(\Delta x)_{\rm min}=\sqrt{e\lambda/2}=1.3861$ ~& $\frac{e}{4}\lambda\kappa_H$ \\
     \hline
    \end{tabular}
  \end{center}
\end{table}
\endgroup

Now, in order to figure out the physical meaning of the all order
GUP corrected entropy, let us rearrange the terms making use of
the surface gravity (\ref{kh-msch}). Then, the modified entropy in
the holographic massive gravity can be rewritten as
\begin{eqnarray}
 \label{massiveEnt}
 S_A &=& S^{{\rm HUP},\tilde{0}} + S^{{\rm GUP},\tilde{0}} + S^{{\rm GUP},\tilde{m}},
 \end{eqnarray}
where
\begin{eqnarray}
 S^{{\rm HUP},\tilde{0}} &=&\pi r^2_+,\\
 \label{gup-massless-ent}
 S^{{\rm GUP},\tilde{0}} &=& \frac{\alpha}{2}+\frac{\beta\lambda}{4r^2_+}, \\
 \label{gup-massive-ent}
 S^{{\rm GUP},\tilde{m}} &=&
            {\cal C}\left[\frac{\alpha}{2}+\frac{\beta\lambda}{4r^2_+}(2+{\cal C})\right]
  +2R\left[2\beta\lambda R+\alpha r_++\frac{\beta\lambda(1+{\cal C})}{r_+}\right].
\end{eqnarray}
Here, $S^{{\rm HUP},\tilde{0}}$ shows the area's law of entropy
with HUP in the holographic massive gravity. Also, $S^{{\rm
GUP},\tilde{0}}$ is the extension of all order GUP corrected
entropy in massless gravity \cite{Kim:2007if} to the holographic
massive gravity. Finally, $S^{{\rm GUP},\tilde{m}}$ has new
contribution to entropy of the holographic massive gravity. In the
massless limit of $R\rightarrow 0$ and ${\cal C}\rightarrow 0$,
the all order GUP corrected entropy is reduced to
 \begin{equation}
 S_A = S^{{\rm HUP},0} + S^{{\rm GUP},0},
 \end{equation}
where
 \begin{eqnarray}
 S^{{\rm HUP},0} &=&\pi (2m)^2,\\
 \label{gup-massless1-ent}
 S^{{\rm GUP},0} &=& \frac{\alpha}{2}+\frac{\beta\lambda}{4(2m)^2}.
 \label{gup-massive1-ent}
 \end{eqnarray}
Note that $2m$ is the radius of the event horizon and $S^{{\rm
GUP},\tilde{m}}=0$ in the massless limit. On the other hand, by
turning off the all order GUP correction to entropy in Eq.
(\ref{massiveEnt}), we have
 \begin{equation}
  S_A = \pi r^2_+,
 \end{equation}
which is the area's law of the Schwarzschild black hole without
the GUP.

\begin{figure}[t!]
   \centering
   \includegraphics{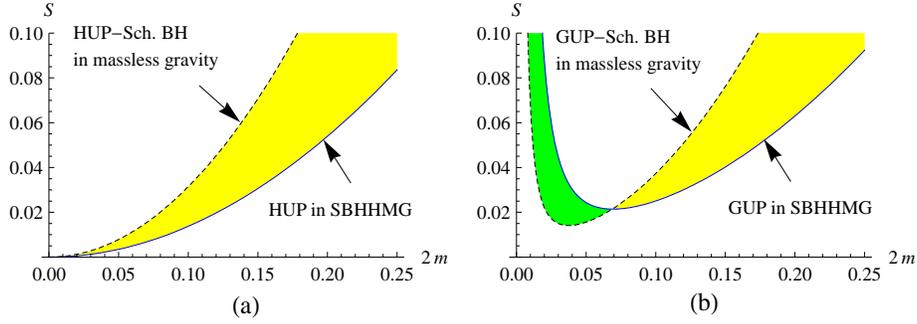}
\caption{Entropy of the Schwarzschild black hole in the massless
and holographic massive gravity with HUP/GUP: (a) the dashed line
is for the Schwarzschild black hole in massless gravity with HUP
($\alpha$, $\beta$, $R$, ${\cal C}=0$) and the solid line is for
the SBHHMG with HUP ($\alpha$, $\beta=0$, $R=0.1$, ${\cal
C}=0.5$), and (b) the dashed line is for the Schwarzschild black
hole in massless gravity with all order corrected GUP ($\alpha$,
$\beta\neq 0$, $R$, ${\cal C}=0$) and the solid line is for the
SBHHMG with all order corrected GUP ($\alpha$, $\beta\neq 0$,
$R=0.1$, ${\cal C}=0.5$). Note that the gaps between the curves
come from the gravitons in the holographic massive gravity. Here,
$\alpha$, $\beta$ are given by Eq. (\ref{values}), respectively.}
\label{fig2}
\end{figure}

In Fig. \ref{fig2}, we have drawn the entropy of the Schwarzschild
black hole in the massless/holographic massive gravity with
HUP/GUP. In Fig. \ref{fig2}(a), the dashed line is for the entropy
of the Schwarzschild black hole in massless gravity with HUP,
which is increasing with $\pi(2m)^2$. When we turn on the massive
gravitons, the ares's law increasing with $\pi (2m)^2$ is changed
to $\pi r^2_+$ since $r_+$ is the event horizon in the holographic
massive gravity. This is represented by the solid line in Fig.
\ref{fig2}(a). On the other hand, the dashed line in Fig.
\ref{fig2}(b) is for the entropy of the Schwarzschild black hole
in massless gravity with all order corrected GUP, while the solid
is for the entropy of the Schwarzschild black hole in holographic
massive gravity with all order corrected GUP. As shown in the
figure, the all order GUP correction makes the entropies divergent
near $m=0$, which was predicted in loop quantum gravity
\cite{Domagala:2004jt,Meissner:2004ju}. Note that the gaps between
the curves come from the gravitons in the holographic massive
gravity.

\begin{figure}[t!]
   \centering
   \includegraphics{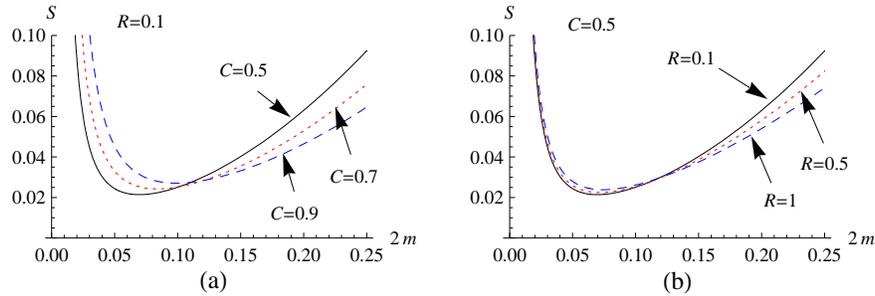}
\caption{Entropy of the SBHHMG: (a) by varying ${\cal
C}=-0.5,~0,~0.5$ with a fixed $R=0.1$ and (b) by varying
$R=0.1,~0.5,~1$ with a fixed ${\cal C}=0.5$.}  \label{fig3}
\end{figure}

In Fig. \ref{fig3}, we have drawn the all order GUP corrected
entropy of the SBHHMG by varying $R$ and ${\cal C}$. Fig.
\ref{fig3}(a) is drawn by varying ${\cal C}$ with a fixed $R$,
while in Fig. \ref{fig3}(b), by varying $R$ with a fixed ${\cal
C}$, respectively. These figures all show results that near $m=0$
all order GUP corrections give dominant effects on the generalized
Bekenstein-Hawking entropy.

\section{Summary and Discussion}


In this paper, we have studied the Hawking temperature and entropy
of a scalar field on the Schwarzschild black hole in the
holographic massive gravity. This is an modification of the
Schwarzschild black hole to have nonzero mass of gravitons. At the
same time, we have studied the effect of a GUP to all orders in
Planck length units to the SBHHMG where a GUP admits a minimal
length at the quantum level and induces quantitative corrections
to the thermodynamic quantities. As a result, we have found the
spherically symmetric Schwarzschild black hole solution with
massive gravitons represented by $R$ and ${\cal C}$ which
contribute linear and constant terms to the solution,
respectively. The allowed event horizons are then classified by
relative magnitudes of $R$ and ${\cal C}$, and even no event
horizon exists when $R<-(1+{\cal C})^2/16m$. We have also obtained
the Schwarzschild-like Hawking temperatures when $R\ge 0$ and
${\cal C}>-1$, and some bizarre ones in the other ranges of $R$
and ${\cal C}$. Thus, one can see that the modified Schwarzschild
black hole provided by the massive gravity and the GUP as a
quantum correction is only possible when $R\ge 0$ and ${\cal
C}>-1$.
On the other hand, we have calculated the statistical entropy by
carefully counting the number of the GUP induced quantum states in
the just vicinity near the horizon. This modification of state
density is due to the GUP where the momentum is a subject of a
modification in the momentum space representation. Comparing with
the previous calculations without massive gravitons
\cite{Li:2002xb,Zhao:2003eu,Liu:2004xh,Kim:2006rx,Yoon:2007aj,Hossenfelder:2006cw,
Hossenfelder:2005ed,Nouicer:2007jg,Nouicer:2007cw,Kim:2007if}
where many authors have expanded the relevant functions only to
the lowest order, we have made use of the substitution method of
integral and finally obtained not only the generalized
Bekenstein-Hawking entropy of the SBHHMG to all orders in the
Planck length units, but also new additional correction terms,
which are proportional to surface gravity. Moreover, we have shown
that the all order GUP corrected entropy obtained in this paper
has the smallest GUP parameter $\lambda$, minimal length, and
thinnest cutoff $\epsilon$, compared with the previous results
\cite{Zhao:2003eu,Liu:2004xh,Kim:2006rx,Yoon:2007aj,Nouicer:2007jg,Nouicer:2007cw,Kim:2007if}.
Furthermore, we have also found that all order GUP corrections to
the Bekenstein-Hawking entropy both in massless and massive
gravities give dominant effects near $m=0$. As a final remark,
while our results are obtained based on the existence of the event
horizons, it would also be interesting to study further the
no-horizon ranges of the massive parameters in Table \ref{evth}
which might have implications on the information loss problem
\cite{Singleton:2010gz,Eune:2014tea} as producing dark stars on a
horizonless background \cite{Kawai:2013mda,Ho:2015vga}.

\acknowledgments{S. T. H. was supported by Basic Science Research
Program through the National Research Foundation of Korea funded
by the Ministry of Education, NRF-2019R1I1A1A01058449. Y. W. K.
was supported by the National Research Foundation of Korea (NRF)
grant funded by the Korea government (MSIT) (No.
2020R1H1A2102242).}

\end{document}